\documentstyle[iopconf1,psfig]{article}
\def\gtwid{\mathrel{\raise.3ex\hbox{$>$\kern-.75em\lower1ex\hbox{$\sim$}}}}
\def\ltwid{\mathrel{\raise.3ex\hbox{$<$\kern-.75em\lower1ex\hbox{$\sim$}}}}

\begin{document}
\title{The cold axion populations}

\author{S. Chang\dag\footnote{Present address: S. Chang, Department of
Physics, Tohoku University,\\ Sendai 980-8578, Japan}, C. Hagmann\ddag\ 
and P. Sikivie\dag\footnote{E-mail:sikivie@phys.ufl.edu}}

\affil{\dag\ Department of Physics, University of Florida,
Gainesville, FL 32601, USA}

\affil{\ddag\ Lawrence Livermore National Laboratory, 
7000 East Avenue, Livermore, CA 94550, USA}

\beginabstract
We give a systematic discussion of the contributions to the cosmological 
energy density in axions from vacuum realignment, string decay and wall 
decay.  We call these the cold axion populations because their kinetic 
energy per particle is at all times much less than the ambient temperature.  
In case there is no inflation after the Peccei-Quinn phase transition, the 
value of the axion mass for which axions contribute the critical energy 
density for closure is estimated to be of order $6 \cdot 10^{-6}$ eV, with 
large uncertainties.  It is emphasized that there are two groups of cold 
axions differing in velocity dispersion by a factor of order $10^3$.
\endabstract

\section{Introduction}

The axion \cite{PQWW,reviews} was postulated approximately twenty years
ago to explain why the strong interactions conserve the discrete symmetries 
P and CP in spite of the fact that the Standard Model of particle interactions
as a whole violates those symmetries.  It is the quasi-Nambu-Goldstone
boson associated with the spontaneous breaking of a $U_{PQ}(1)$ symmetry
which Peccei and Quinn postulated.  At zero temperature the axion mass is
given by:
\begin{equation}
m_a\simeq 6 \cdot 10^{-6} \mbox{eV} \cdot N \cdot \left(\frac{10^{12}
\mbox{GeV}} {v_a} \right) 
\label{1.1}
\end{equation}
where $v_a$ is the magnitude of the vacuum expectation value that breaks   
$U_{PQ}(1)$ and $N$ is a strictly positive integer that describes the
color anomaly of $U_{PQ}(1)$.  The combination of parameters 
$f_a \equiv v_a/N$ is usually called the axion decay constant.  Axion models 
have $N$ degenerate vacua \cite{degen,reviews}.  Searches for the axion in 
high energy and nuclear physics experiments have only produced negative 
results.  By combining the constraints from these experiments with those 
from astrophysics \cite{reviews,astro}, one obtains the bound: 
$m_a \ltwid 10^{-2}$ eV.

The axion owes its mass to non-perturbative QCD effects.  In the very
early universe, at temperatures high compared to the QCD scale, these
effects are suppressed \cite{highT} and the axion mass is negligible. The
axion mass turns on when the temperature approaches the QCD scale and
increases till it reaches the value given in Eq.(\ref{1.1}) which is valid 
below the QCD scale.  There is a critical time $t_1$, defined by 
$m_a(t_1) t_1 = 1$, when the axion mass effectively turns on \cite{vacmis}. 
The corresponding temperature $T_1 \simeq 1$ GeV.

The implications of the existence of an axion for the history of the
early universe may be briefly described as follows.  At a temperature of   
order $v_a$, a phase transition occurs in which the $U_{PQ}(1)$ symmetry   
becomes spontaneously broken.  This is called the PQ phase transition.  
At that time axion strings appear as topological defects.  One must 
distinguish two cases: 1) inflation occurs with reheat temperature higher
than the PQ transition temperature (equivalently, for the purposes of this
paper, inflation does not occur at all) or 2) inflation occurs with reheat 
temperature less than the PQ transition temperature.

In case 2 the axion field gets homogenized by inflation and the axion
strings are `blown away'.  When the axion mass turns on at $t_1$, the      
axion field starts to oscillate.  The amplitude of this oscillation
is determined by how far from zero the axion field is when the axion
mass turns on.  The axion field oscillations do not dissipate into
other forms of energy and hence contribute to the cosmological
energy density today \cite{vacmis}.  Such a contribution is called of
``vacuum realignment".  Note that the vacuum realignment contribution
may be accidentally suppressed in case 2 because the axion field, which
has been homogenized by inflation, may happen to lie close to zero.

In case 1 the axion strings radiate axions \cite{Davis,Harari} from the 
time of the PQ transition till $t_1$ when the axion mass turns on.   At 
$t_1$ each string becomes the boundary of $N$ domain walls.  If $N=1$, 
the network of walls bounded by strings is unstable \cite{Vil,Paris} and
decays away.  If $N>1$ there is a domain wall problem \cite{degen} because
axion domain walls end up dominating the energy density, resulting in a
universe very different from the one observed today.  There is a way
to avoid the domain wall problem by introducing an interaction which 
slightly lowers one of the $N$ vacua with respect to the others.  In that 
case, the lowest vacuum takes over after some time and the domain walls 
disappear.  There is little room in parameter space for that to happen 
and we will not consider this possibility further here.  A detailed 
discussion is given in Ref.\cite{axwall}.  Here we assume $N=1$.

In case 1 there are three contributions to the axion cosmological
energy density.  One contribution \cite{Davis,Harari,stringA,Hagmann,hcp}
is from axions that were radiated by axion strings before $t_1$; let us
call it the string decay contribution.  A second contribution is from
axions that were produced in the decay of walls bounded by strings after
$t_1$ \cite{Hagmann,Lyth,Nagasawa,axwall}; call it the contribution from
wall decay.  A third contribution is from vacuum realignment
\cite{vacmis}.  To convince oneself that there is a vacuum realignment
contribution distinct from the other two, consider a region of the
universe which happens to be free of strings and domain walls.  In such a
region the axion field is generally different from zero, even though no
strings or walls are present. After $t_1$, the axion field oscillates
implying a contribution to the energy density which is neither from string
decay nor wall decay.  Since the axion field oscillations caused by vacuum
realignment, string decay and wall decay are mutually incoherent, the
three contributions to the energy density should simply be added to each
other \cite{Hagmann}.   

We will see below that the axions from vacuum realignment, string decay and 
wall decay are {\it cold}, i.e. their effective temperature is much smaller 
than the temperature of the ambient photons.  In addition to cold axions,  
there is a thermal axion population whose properties are derived by the usual
application of statistical mechanics to the early, high temperature
universe \cite{KT}.  The thermal axions have an effective temperature 
of order that of the ambient photons.  The contribution of thermal axions
to the cosmological energy density is subdominant if $m_a
\ltwid 10^{-2}$ eV.  We concern ourselves only with cold axions here.

The next three sections discuss the contributions to the cosmological axion
energy density from:\\
1. vacuum realignment\\
\indent\hspace{.5cm}a. ``zero momentum'' mode\\
\indent\hspace{.5cm}b. higher momentum modes\\
2. axion string decay\\
3. axion domain wall decay.\\
The basis for distinguishing the contribution from vacuum realignment
\cite{vacmis} from the other two is that it is present for any
quasi-Nambu-Goldstone field regardless of the topological structures
associated with that field.  Contributions 1a and lb are distinguished by
whether the misalignment of the axion field from the CP conserving
direction is in modes of wavelength larger (1a) or shorter (1b) than the
horizon size $t_1$ at the moment the axion mass turns on.  Contribution 2
is from the decay of axion strings before the QCD phase transition.  
Contribution 3 is from the decay of axion domain walls bounded by strings
after the QCD phase transition.  In case 1, all contributions are
present.  In case 2, contributions 1b, 2 and 3 are absent or at any rate
much suppressed; contribution 1a is present but, as was already mentioned,
it may be accidentally small.

In section 5, we estimate the total axion cosmological energy density.  In
section 6, we point out that there are two groups of cold axions (pop.~I and 
pop.~II) differing in velocity dispersion by a factor of order $10^3$.

\section{Vacuum realignment}

Let us track the axion field in case 1 from the PQ phase transition,
when $U_{PQ}(1)$ gets spontaneously broken and the axion appears as a
Nambu-Goldstone boson, to the QCD phase transition when the axion acquires
mass.  For clarity of presentation we consider in this section a large
region which happens to be free of axion strings.  Although such a region
is rare in practice it may exist in principle.  In it, the contribution to
the cosmological axion energy density from string and domain wall decay
vanishes but that from vacuum realignment does not.  In more typical
regions where strings are present, the contributions from string and
domain wall decay should simply be added to that from vacuum realignment
\cite{Hagmann}.

In the radiation dominated era under consideration, the space-time metric
is given by:
\begin{equation}
- ds^2 = - dt^2 + R^2 (t)~d\vec x\cdot d\vec x
\label{3.1}
\end{equation}
where $t$ is the age of the universe, $\vec x$ are co-moving coordinates 
and $R(t) \sim \sqrt{t}$ is the scale factor.  The axion field $a(x)$
satisfies:
\begin{eqnarray}
\left( D_\mu D^\mu + m_a^2 (t)\right) a(x)
= \left(\partial_t^2 + 3 \frac{\dot R}{R} \partial_t - \frac{1}{R^2}
\nabla_x^2 + m_a^2(t)\right) a(x) = 0
\label{3.2}
\end{eqnarray}
where $m_a (t) = m_a (T(t))$ is the time-dependent axion mass.  We will
see that the axion mass is unimportant as long as $m_a(t) \ll 1/t$, a
condition which is satisfied throughout except at the very end when $t$
approaches $t_1$.  The solution of Eq.~(\ref{3.2}) is a linear
superposition of eigenmodes with definite co-moving wavevector $\vec k$:
\begin{equation}
a (\vec x, t) = \int d^3 k~~a(\vec k, t)~e^{i\vec k\cdot\vec
x}
\label{3.3}
\end{equation}
where the $a(\vec k, t)$ satisfy:
\begin{equation}
\left( \partial_t^2 + {3\over 2t} \partial_t + {k^2\over R^2} + m_a^2   
(t)\right) a(\vec k, t) = 0\ .
\label{3.4}
\end{equation}  
Eqs.~(\ref{3.1}) and (\ref{3.3}) show that the wavelength $\lambda (t)
= {2\pi\over k} R(t)$ of each mode is stretched by the Hubble expansion.
There are two qualitatively different regimes in the evolution of each
mode depending on whether its wavelength is outside $(\lambda (t) > t)$ or
inside $(\lambda (t) < t)$ the horizon.

For $\lambda (t) \gg t$, only the first two terms in Eq.~(\ref{3.4})
are important and the most general solution is:
\begin{equation}
a(\vec k, t) = a_0 (\vec k) + a_{-1/2} (\vec k) t^{-1/2}\ .
\label{3.5}
\end{equation}
Thus, for wavelengths larger than the horizon, each mode goes to 
a constant; the axion field is so-called ``frozen by causality''.

For $\lambda (t) \ll t$, the first three terms in Eq.~(\ref{3.4}) are
important.  Let $a(\vec k, t) = R^{-{3 \over 2}}(t) \psi(\vec k, t)$.
Then
\begin{equation}
\left( \partial_t^2 + \omega^2(t)\right) \psi(\vec k, t) = 0 \,
\end{equation}
where
\begin{equation}
\omega^2(t) = m_a^2(t) + {k^2\over R^2 (t)} + {3\over 16t^2} \simeq
{k^2\over R^2(t)}\ .
\end{equation}
Since $\dot\omega \ll \omega^2$, this regime is characterized by
the adiabatic invariant $\psi_0^2(\vec k,t)\omega(t)$, where
$\psi_0(\vec k,t)$  is the oscillation amplitude of $\psi(\vec k,t)$.
Hence the most general solution is:
\begin{equation}
a (\vec k, t) = {A\over R(t)} \cos \left( \int^t dt^\prime \omega
(t^\prime) \right)~~~~~\ .
\label{3.6}
\end{equation}
The energy density and the number density behave respectively as
$\rho_{a,\vec k} \sim {A^2 w^2\over R^2(t)} \sim{1\over R^4 (t)}$
and $n_{a,\vec k} \sim {1\over \omega} \rho_{a,\vec k}
\sim {1\over R^3 (t)},$ indicating that the number of axions in each 
mode is conserved.  This is as expected because the expansion of the
universe is adiabatic for $\lambda (t) t \ll 1$.

Let us define ${dn_a\over dw} (\omega, t)$ to be the number density, in
physical and frequency space, of axions with wavelength
$\lambda=\frac{2\pi}{\omega}$, for $\omega >t^{-1}$.
The axion number density in physical space is thus:
\begin{equation}
n_a (t) = \int_{t^{-1}} d\omega~{dn_a\over dw} (\omega, t)\ ,
\label{3.8}
\end{equation}
whereas the axion energy density is:
\begin{equation}
\rho_a (t) = \int_{t^{-1}} d\omega {dn_a\over d\omega} (\omega, t)
\omega\ .
\label{3.9}
\end{equation}
Under the Hubble expansion axion energies redshift according to
$\omega^\prime = \omega {R\over R^\prime}$, and volume elements expand
according to
$\Delta V^\prime = \Delta V\left({R^\prime\over R}\right)^3$, whereas
the number of axions is conserved mode by mode. Hence:
\begin{equation}
{dn_a\over d\omega} (\omega, t) = \left({R^\prime\over R}\right)^2
{dn_a\over d\omega} (\omega {R\over R^\prime}, t^\prime)\ .
\label{3.10}
\end{equation}
Moreover, the size of ${dn_a\over d\omega}$ for $\omega \sim {1\over t}$
is determined in order of magnitude by the fact that the axion field  
typically varies by $f_a$ from one horizon to the next.  Thus:
\begin{equation}
\left.\omega {dn_a\over d\omega} (\omega, t) \Delta\omega \right|_{\omega
\sim\Delta\omega\sim \frac{1}{t}}
\sim {dn_a\over d\omega}
\left({1\over t}, t\right) \left({1\over
t}\right)^2 \sim {1\over 2} (\vec\nabla a)^2 \sim {1\over 2} {f_a^2\over
t^2}\ .
\label{3.11}
\end{equation}
{}From Eqs.~(\ref{3.10}) and (\ref{3.11}), and $R \sim \sqrt{t}$, we have:
\begin{equation}
{dn_a\over d\omega} (\omega, t) \sim {f_a^2\over 2t^2\omega^2}\ .
\label{3.12}
\end{equation}
Eq.~(\ref{3.12}) holds until the moment the axion acquires mass during the
QCD
phase transition.  The critical time is when $m_a(t)$ is of order
$t^{-1}$.  Let us define $t_1$:
\begin{equation}
m_a (t_1) t_1 = 1 \ .
\label{3.13}
\end{equation}  
$m_a(T)$ was obtained \cite{vacmis} from a calculation of the effects of
QCD
instantons at high temperature \cite{highT}:
\begin{equation}
m_a(T) \simeq 4 \cdot 10^{-9} {\rm eV} \left(\frac{10^{12}\rm
GeV}{f_a}\right)
\left(\frac{\rm GeV}{T}\right)^4
\end{equation}
when $T$ is near $1$ GeV. The relation between $T$ and $t$ follows as
usual
from
\begin{equation}
H^2=\left(\frac{1}{2t}\right)^2=\frac{8\pi G}{3} \rho = \frac{8\pi G}{3}
\cdot \frac{\pi^2}{30} {\cal N} T^4
\end{equation}
where ${\cal N}$ is the effective number of thermal degrees of freedom.
${\cal N}$ is changing near 1 GeV temperature from a value near 60, valid
above
the quark-hadron phase transition, to a value of order 30 below that
transition.
Using ${\cal N} \simeq 60$, one has
\begin{equation}
m_a(t) \simeq 0.7~10^{20} \frac{1}{\rm sec}\left(\frac{t}{\rm
sec}\right)^2
\left(\frac{10^{12}\rm GeV}{f_a}\right)  \ ,
\label{3.16n}
\end{equation}
which implies:
\begin{equation}
t_1 \simeq 2\cdot 10^{-7} \mbox{sec}
\left({f_a\over 10^{12} \mbox{GeV}}\right)^{1/3} \ .
\label{3.14}
\end{equation}  
The corresponding temperature is:
\begin{equation}
T_1 \simeq 1 \mbox{GeV}
\left({10^{12} \mbox{GeV}\over f_a}\right)^{1/6}.
\label{3.15}  
\end{equation}  
We will make the usual assumption \cite{vacmis} that the changes in the
axion mass are adiabatic [i.e. ${1\over m_a(t)} {dm_a\over dt} \ll
m_a(t)$] after $t_1$ and that the axion to entropy ratio is constant from
$t_1$ till the present.  Various ways in which this assumption may be
violated are discussed in the papers of ref.~\cite{entrop}.  We also
assume that the axions do not convert to some other light axion-like
particles as discussed in ref.~\cite{Hill}.   

The above analysis neglects the non-linear terms associated with the
self-couplings of the axion.  In general, such non-linear terms may cause
the higher momentum modes to become populated due to spinodal
instabilities and parametric resonance.  However, in the case of the axion
field, such effects are negligible \cite{vacmis,Singh}.

We are now ready to discuss the vacuum realignment contributions to the
cosmological axion energy density.

\subsection{Zero momentum mode}

This contribution is due to the fact that, at time $t_1$, the axion field 
averaged over distances less than $t_1$ has in each horizon volume a random 
value between $-\pi f_a$ and $+\pi f_a$, whereas the CP conserving, and 
minimum energy density, value is $a=0$.  In case 1 the average energy 
density in this ``zero momentum mode" at time $t_1$ is of order:
\begin{equation}
\rho_a^{vac,0} (t_1) \sim {1\over 2} m_a^2 (t_1) f_a^2 
\langle \alpha^2(t_1) \rangle \sim {1\over 2} m_a^2 (t_1) f_a^2
~~~{\rm (case~1)} \ ,
\label{3.16}
\end{equation}
Here $\alpha(t_1) \equiv a(t_1)/f_a$ is the initial misalignment angle.
The average $\langle \alpha^2(t_1) \rangle$ is of order one because
$\alpha(t_1)$ is randomly and independently chosen in each horizon 
volume. The corresponding average axion number density is:
\begin{eqnarray}
n_a^{vac,0} (t_1) = {1\over m_a(t_1)} \rho_a^{vac,0}(t_1) \sim 
{f_a^2\over 2 t_1}~~~{\rm (case~1)} \ .
\label{3.17}
\end{eqnarray}
Since $m_a(t)$ is assumed to change adiabatically after $t_1$, the number
of axions is conserved after that time.  Hence:
\begin{equation}
n_a^{vac,0} (t) \sim {1\over 2} {f_a^2\over t_1} \left({R_1\over
R}\right)^3\ ~~~~~~~~~~{\rm (case~1)}.
\label{3.18}
\end{equation}
The axions in this population are non-relativistic and contribute $m_a$
each to the energy.  Their average kinetic energy is at all times much 
less than the ambient temperature.

In case 2 the axion field has been homogenized by inflation and has 
everywhere the same initial value $a(t_1) = f_a \alpha(t_1)$.  Hence 
\begin{equation}
n_a^{vac,0} (t) \sim {1\over 2} {f_a^2\over t_1} \left({R_1\over
R}\right)^3 \alpha^2(t_1)  ~~~~{\rm (case~2)}  \ .
\label{3.182}
\end{equation}
The cold axion density is suppressed if $\alpha(t_1)$ happens to be small.  
The probability that it be suppressed by the factor $x$ is of order 
$\sqrt{x}$.

\subsection{Higher momentum modes}
This contribution is due to the fact that the axion field has wiggles
about its average value inside each horizon volume.  The axion number
density and spectrum associated with these wiggles is given in
Eq.~(\ref{3.12}) for case 1.  Integrating over $\omega > t^{-1}$, 
we find:
\begin{equation}
n_a^{vac,1} (t) = n_a^{vac,1} (t_1) \left({R_1\over R}\right)^3 \sim   
{1\over 2} {f_a^2\over t_1} \left({R_1\over R}\right)^3~~~~
{\rm(case~1)}
\label{3.19}
\end{equation}
for the contribution from vacuum realignment associated with higher
momentum modes.  The bulk of these axions are non-relativistic after 
time $t_1$ and hence contribute $m_a$ each to the energy.  Note
that $n_a^{vac,0} (t)$ and $n_a^{vac,1} (t)$ have the same order
of magnitude.

In case 2, the field is homogenized by inflation and hence 
$n_a^{vac,1}$ is very much suppressed.  It does not vanish completely 
because of quantum-mechanical fluctuations in the axion field during the 
inflationary epoch \cite{infl}.  These fluctuations are important 
if $f_a$ approaches the Planck scale.  However, for $f_a \sim 10^{12}$ GeV,
the resulting $n_a^{vac,1}$ is very small.

\section{String decay}

In the early universe, the axion is essentially massless from the PQ phase
transition at temperature of order $v_a$ to the QCD phase transition at
temperature of order $1$ GeV. During that epoch, axion strings are present
as topological defects, assuming case 1 as we do henceforth.  The following 
model describes the relevant dynamics:
\begin{equation}
{\cal L} = \frac{1}{2} \partial_{\mu} \phi^\dagger \partial^{\mu} \phi
-\frac{\lambda}{4}(\phi^\dagger\phi-v_a^2)^2 
\label{n26}
\end{equation}
where $\phi=\phi_1+ i\phi_2$ is a complex scalar field.  The axion 
field $a(x)$ is $f_a$ times the phase of $\phi(x)$.  The classical 
solution
\begin{equation}
\phi(\vec{x})=v_a f(\rho) e^{i\theta}
\end{equation}
describes a static, straight global string lying on the $\hat{z}$-axis.
Here $(z,\rho,\theta)$ are cylindrical coordinates and $f(\rho)$ is a   
function which goes from zero at $\rho=0$ to one at $\rho=\infty$ over a
distance scale of order $\delta \equiv (\sqrt{\lambda} v_a)^{-1}$.
$\delta$ is called the core size.  $f(\rho)$ may be determined by solving
the non-linear field equations derived from Eq.~(\ref{n26}). The energy per 
unit length of the global string is
\begin{equation}
\mu \simeq  2\pi \int^L_\delta \rho d\rho \frac{1}{2} |\vec{\nabla}\phi|^2
= \pi v_a^2 \ln (\sqrt{\lambda}v_aL)
\label{2.3}   
\end{equation}
where $L$ is a long-distance cutoff. Eq.~(\ref{2.3}) neglects the energy per
unit length, of order $v_a^2$, associated with the core of the string.
For a network of strings with random directions, as would be present in
the early universe, the infra-red cutoff $L$ is of order the distance
between strings.

At first the strings are stuck in the plasma and are stretched by the
Hubble expansion.  However with time the plasma becomes more dilute and
below temperature \cite{Harari}
\begin{equation}
T_* \sim 2~10^7 \mbox{GeV} \left({v_a \over 10^{12} \mbox{GeV}}\right)^2
\label{2.7}
\end{equation}
the strings move freely.  String loops of size smaller than the horizon 
decay rapidly into axions.  The strings that traverse the horizon, called
{\it long} strings, intersect each other with reconnection thus producing 
sub-horizon sized loops which decay efficiently.  As a result there is 
approximately one long string per horizon from temperature $T_*$ to 
temperature $T_1\simeq 1$ GeV when the axion acquires mass.  

One long axion string per horizon implies the energy density:
\begin{equation}
\rho_{str} (t) \sim {\pi v_a^2\over t^2} \ln (t v_a)\ .
\label{3.21}
\end{equation}
We are interested in the {\it number} density $n_a^{str}(t)$ of axions 
produced in the decay of axion strings because, as we will soon see, these 
axions become non-relativistic soon after time $t_1$ and hence contribute 
each $m_a$ to the energy.  The equations governing $n_a^{str}(t)$ are:
\begin{equation}
{d\rho_{str}\over dt} = -2 H\rho_{str} - {d\rho_{str\rightarrow
a}\over dt}
\label{3.22}
\end{equation}
\begin{equation}
{dn_a^{str}\over dt} = - 3H n_a^{str} + {1\over \omega (t)}
{d\rho_{str\rightarrow a}\over dt}
\label{3.23}
\end{equation}
where $\omega (t)$ is defined by:
\begin{equation}
{1\over \omega (t)} = {1\over {d\rho_{str\rightarrow a}\over dt}} \int
{d\omega\over \omega} {d\rho_{str\rightarrow a}\over dt~d\omega}\ .
\label{3.24}  
\end{equation}  
In Eqs.~(\ref{3.22}-\ref{3.24}), ${d\rho_{str\rightarrow a}\over dt}$
is the rate at which energy density gets converted from strings into
axions at time $t$, whereas ${d\rho_{str\rightarrow a}\over dt~d\omega}$
is the spectrum of axions thus emitted.  The term $-2H \rho_{str} = + H
\rho_{str} - 3 H\rho_{str}$ in Eq.~(\ref{3.22}) takes account of the fact
that the Hubble expansion both stretches $(+ H\rho_{str})$ and dilutes
$(-3 H\rho_{str})$ long strings.  To obtain $n_a^{str} (t)$ from
Eqs.~(\ref{3.21}-\ref{3.22}), we must know what $\omega (t)$ characterizes
the spectrum ${d^2\rho_{str\rightarrow a}\over dt~d\omega}$ of axions
radiated by cosmic axion strings at time $t$.

The axions that are radiated at time $t$ are emitted by strings which
are bent over a distance scale of order $t$ and are relaxing to lower
energy configurations.  The main source is closed loops of size $t$. Two
views have been put forth concerning the motion and the radiation spectrum
of such loops.  One view \cite{Davis,stringA} is that such a loop oscillates 
many times, with period of order $t$, before it has released its excess
energy and that the spectrum of radiated axions is concentrated near
${2\pi\over t}$.  Let us call this scenario $A$.  A second view 
\cite{Harari,Hagmann} is that the loop releases its excess energy very
quickly and that the spectrum of radiated axions
${d\rho_{str\rightarrow a}\over dt~d\omega} \sim {1\over \omega}$
with a high energy cutoff of order $v_a$ and a low energy cutoff of order
${2\pi\over t}$.  Let us call this scenario $B$.  In scenario $A$ one
has $\omega (t) \sim {2\pi\over t}$ and hence, for $t < t_1$:
\begin{equation}
n_a^{str,A} (t) \sim {v_a^2\over t} \ln (v_a t) \ ,
\label{3.25}
\end{equation}
whereas in $B$ one has $\omega (t) \sim {2\pi\over t} \ln (v_a t)$ and
hence:
\begin{equation}
n_a^{str,B} (t) \sim {v_a^2\over t}\ .
\label{3.26}
\end{equation}
We have carried out computer simulations \cite{Hagmann,hcp} of the motion
and decay of axion strings with the purpose of deciding between the two
possibilities.  We define the quantity:  
\begin{equation}
N_{ax} \equiv \int dk {dE\over dk} {1\over k}\ ,
\label{3.27}
\end{equation}
and compute it as a function of time while the strings decay into axions.  
We have
\begin{equation}
{1 \over \omega(t)} \simeq r {2\pi \over t} {1 \over \ln(v_a t)}
\end{equation}
where $r \equiv N_{ax}^{\rm final}/N_{ax}^{\rm initial}$ is the factor by 
which $N_{ax}$ increases during the decay of a string loop.  Integrating
Eqs.~(\ref{3.22}) and (\ref{3.23}), one finds:
\begin{equation}
n_a^{str} (t) \sim r {v_a^2\over t_1} \left({R_1\over R}\right)^3\ .   
\label{3.28}
\end{equation}
The string decay contribution is of order $r$ times the contribution from 
misalignment.  In scenario A, $r$ is of the order $\ln(t_1/\delta)\simeq 70$, 
whereas in scenario B, $r$ is of order one.

We have performed simulations of two loop geometries: circular loops 
initially at rest, and noncircular loops with angular momentum. The initial 
configurations are set up on large ($\sim 10^7$ points) Cartesian grids,
\begin{figure}
\psfig{figure=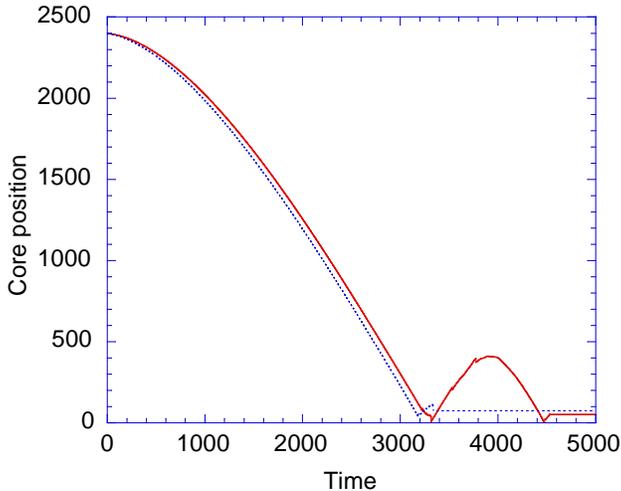,height=3.0in,bbllx=0in,bblly=2in,bburx=9in,bbury=8in}
\caption{\small{Core position of collapsing loop versus time for $\lambda
= 0.001$ (dotted line) and $\lambda  = 0.004$ (solid line).  The lattice
size is $4096\times4096$.}}
\end{figure}
and then time-evolved using the finite difference equations which follow 
from Eq.~(\ref{n26}).  FFT spectrum analysis of the kinetic and gradient 
energies during the collapse yields $N_{ax}(t)$. 

\subsection{Computer simulations of circular loops}

Because of azimuthal symmetry, circular loops can be studied in $r-z$ space. 
By mirror symmetry, the problem can be further reduced to one quarter-plane. 
The static axion field far from the string core is 
\begin{equation}
a(r,z) = \frac{f_a}{2} \Omega(r,z)
\label{n39}
\end{equation}
in the infinite volume limit, where $\Omega$ is the solid angle subtended by 
the loop. We use as initial configuration the outcome of a relaxation
routine starting with Eq.~(\ref{n39}) outside the core and
$\phi(\rho)\approx 0.58(\rho/\delta)\exp(i\theta)$ within the core. Here, 
$\rho$ is the distance to the string, and $\theta$ is the winding angle. The 
relaxation and the subsequent dynamical evolution are done with reflective 
boundary conditions. A step size $dt$ = 0.2 was used for the time evolution 
and the total energy was conserved to better than 1 \%.
\begin{figure}[t]
\psfig{figure=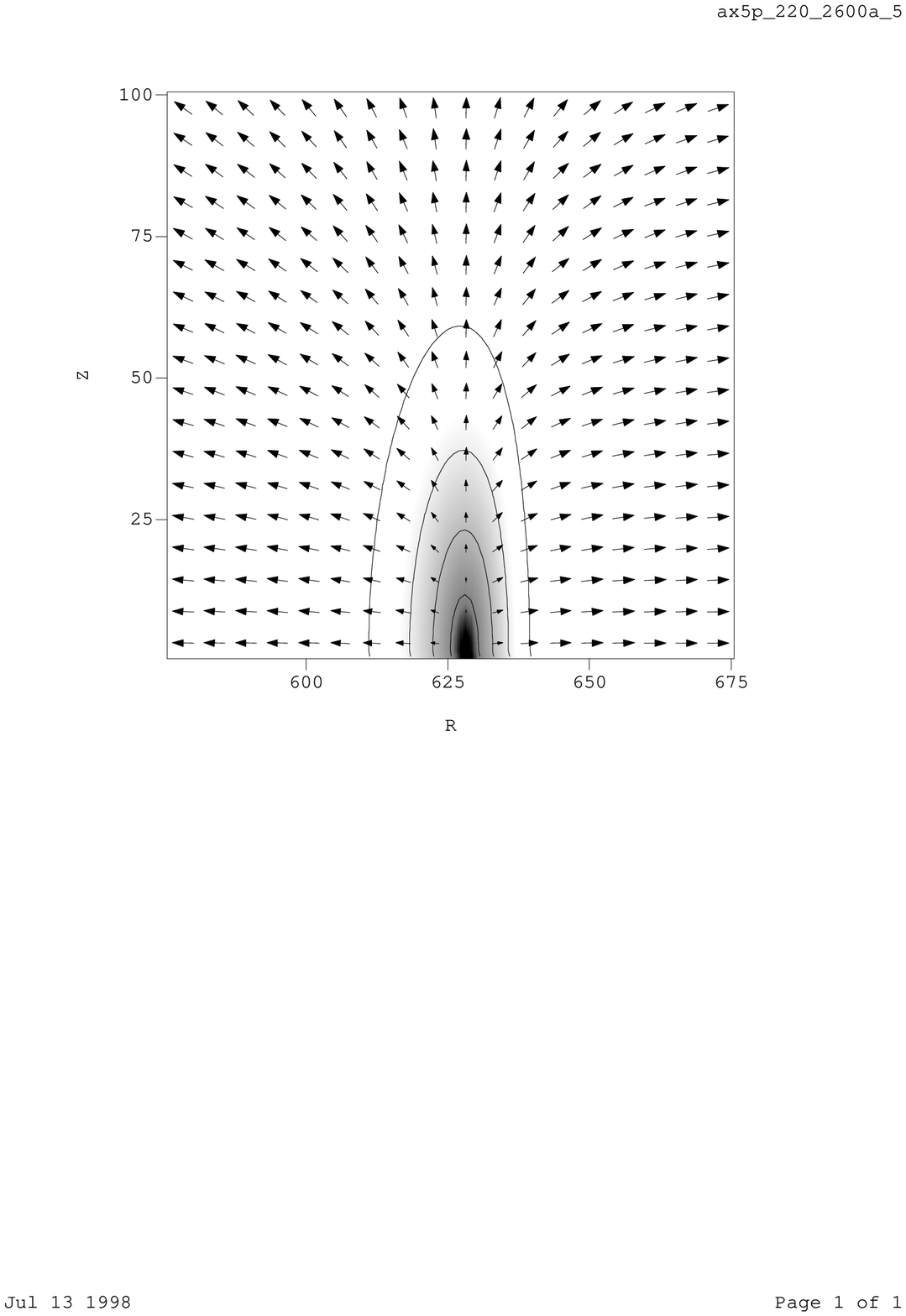,height=3.0in,bbllx=0in,bblly=3in,bburx=9in,bbury=10in}
\caption{\small{Intensity plot of potential energy (contours represent
constant potential energy) in vicinity of string core for $R_0 = 2400, 
\lambda = 0.001$ at $t = 2600$.  The Lorentz factor is about 4.  The 
arrows represent the axion field.}}
\end{figure}
In general the loops collapsed at nearly the speed of light without a rebound. 
For a small range of parameters, $80 < R_0/\delta < 190$, where $R_0$ is the 
initial loop radius, we noticed a small bounce as show in Figure 1. There is 
a substantial Lorentz contraction of the string core as it collapses (see
Figure 2). A lattice effect became evident when the reduced core size becomes 
comparable to the lattice spacing. This consists of a ``scraping'' of the 
string core on the underlying grid, during which the kinetic energy of the 
string gets dissipated into high frequency axion radiation.  We always choose 
$\lambda$ small enough to avoid this phenomenon.

Spectrum analysis of the fields was performed by expanding the 
gradient and kinetic energies as
\begin{eqnarray}
\nabla_z\phi=\sum_{mn}c_{mn} J_0(k_mr)\,\sin(k_nz) \nonumber\\
\nabla_r\phi=\sum_{mn}c_{mn} J_1(k_mr)\,\cos(k_nz) \nonumber\\
\dot{\phi}=  \sum_{mn}c_{mn} J_0(k_mr)\,\cos(k_nz)
\end{eqnarray}
with the boundary conditions $J_1(k_mR_{\rm max})=0$ and 
${\rm sin}(k_nZ_{\rm max})=0$. The dispersion relationship for axions is 
given by $\omega_{mn}=\sqrt{k_m^2+k_n^2}$.  Figure 3 shows the power 
spectrum $dE/d\ln k$ displayed in ${\rm ln}k$ bins of width 
$\Delta \ln k=0.5$, at $t=0$ and after the collapse at $t=3000$. At both 
times, the spectrum 
\begin{figure}[t]
\psfig{figure=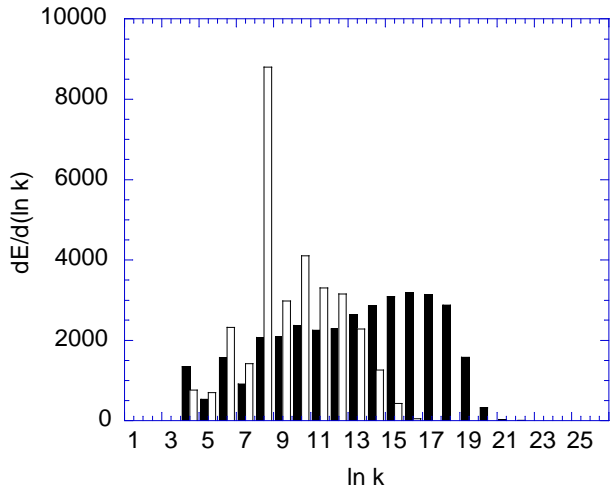,height=3.0in,bbllx=0in,bblly=2in,bburx=9in,bbury=8in}
\caption{\small{Energy spectrum of collapsing loop for $R_0 = 2400$ and 
$\lambda = 0.001$.  The white (black) histogram represents the spectrum 
at $t = 0 (t = 3000).$  The increased high frequency cutoff of the final 
spectrum is due to the Lorentz contraction of the core.}}
\end{figure}
exhibits an almost flat plateau, consistent with $dE/dk\propto 1/k$.  The 
high frequency cutoff of the spectrum is increased however after the collapse 
and is associated with the Lorentz contraction of the core.  The evolution of
$N_{ax}=\sum_{mn}(E_{mn}/k_{mn})$ during the loop 
\begin{figure}[t]
\psfig{figure=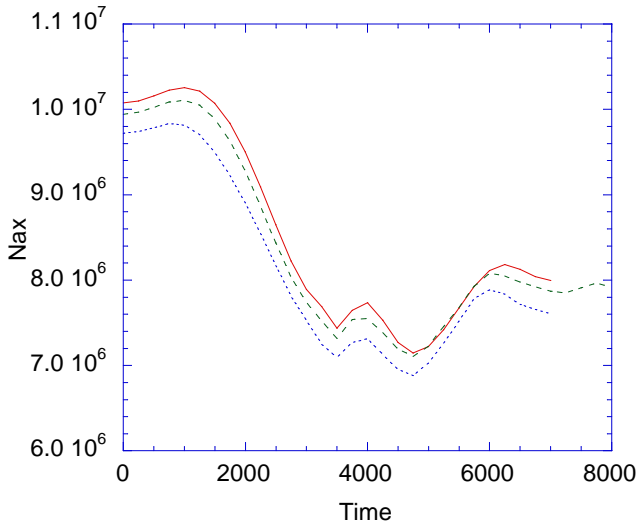,height=3.0in,bbllx=0in,bblly=2in,bburx=9in,bbury=8in}
\caption{\small{$N_{ax}$ of a circular loop as a function of time for 
$R_0 = 2400$, and $\lambda = 0.004$ (solid line), $\lambda = 0.001$ 
(dashed line), and $\lambda = 0.00025$ (dotted line).}}
\end{figure}
collapse was studied for various values of $R_0/\delta$. We observe a
marked decrease of $N_{ax}$ by $\sim 20 \%$ during the collapse roughly 
independent of $R_0/\delta$. 

\subsection{Computer simulations of rotating loops}

There exists a family of nonintersecting (``Kibble-Turok'') loops 
\cite{kibb82,turo84,burd85}, which have been studied in the context of gauge 
strings. They are solutions of the Nambu-Goto equations of motion and have 
non-zero angular momentum.  Intercommuting (self-intersection with 
reconnection) causes the loop sizes to shrink, and hence the average energy 
of radiated axions to increase and hence $N_{ax}$ to decrease. Intercommuting 
favors scenario B for these reasons.  We picked the Kibble-Turok configuration 
as an initial condition to avoid intercommuting as much as possible and give 
scenario A the best possible chance to get realized. A common loop 
parametrization is given by \cite{turo84}
\begin{eqnarray}
x(\sigma,t)&=&\frac{R}{2}\left((1-\alpha)\,\sin \sigma_-
           +\frac{1}{3}\alpha\,\sin 3\sigma_-
           +\sin\sigma_+ \right) \nonumber\\
y(\sigma,t)&=&\frac{R}{2}\left(-(1-\alpha)\,\cos \sigma_-
           -\frac{1}{3}\alpha\,\sin 3\sigma_-
           -\cos \psi\,\cos \sigma_+ \right) \nonumber\\
z(\sigma,t)&=&\frac{R}{2}\left(-2\sqrt{\alpha(1-\alpha)}\,\cos \sigma_-
           -\sin \psi\,\cos \sigma_+ \right) 
\end{eqnarray}
where $\sigma_\pm=(\sigma\pm t)/R$, and $\sigma\in(0,2\pi R)$ is the length 
along the loop. For a significant subset of the free parameters $\alpha\in
(0,1)$, $\psi\in (-\pi,\pi)$, the loop never self-intersects. A noteworthy 
feature in the case of gauge strings is the periodic appearance of cusps, 
where the string velocity momentarily reaches the speed of light. 

We performed numerous simulations of rotating loops on a 3D $(256^3)$ lattice 
with periodic boundary conditions. Standard Fourier techniques were used  
\begin{figure}[t]
\psfig{figure=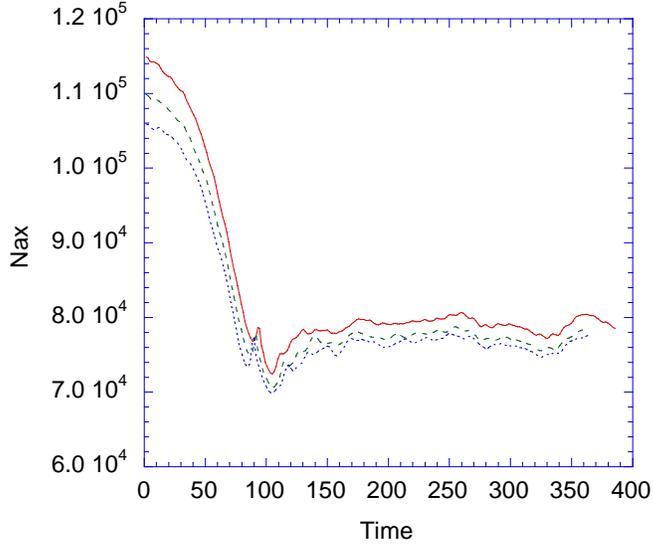,height=3.0in,bbllx=0in,bblly=2in,bburx=9in,bbury=8in}
\caption{\small{$N_{ax}$ of non-intersecting ("Kibble-Turok") loops as a 
function of time for $\alpha = 0.01, \phi = 0$, and $\lambda = 0.2$ (solid 
line), $\lambda = 0.1$ (dashed line), and $\lambda = 0.0625$ (dotted line).  
The lattice size is $256^3$ and $R = 72$.}}
\end{figure}
for the spectrum analysis, and $N_{ax}$ was computed as a function of time 
using the dispersion relationship 
\begin{equation}
\omega_{mnp}=\sqrt{2(3-\cos k_m-\cos k_n-\cos k_p)}
\end{equation}
Figure 5 shows $N_{ax}$ for various core sizes and constant $\alpha, \psi$. 
The behavior is very similar to that of a non-rotating circular loop with a 
\begin{figure}
\psfig{figure=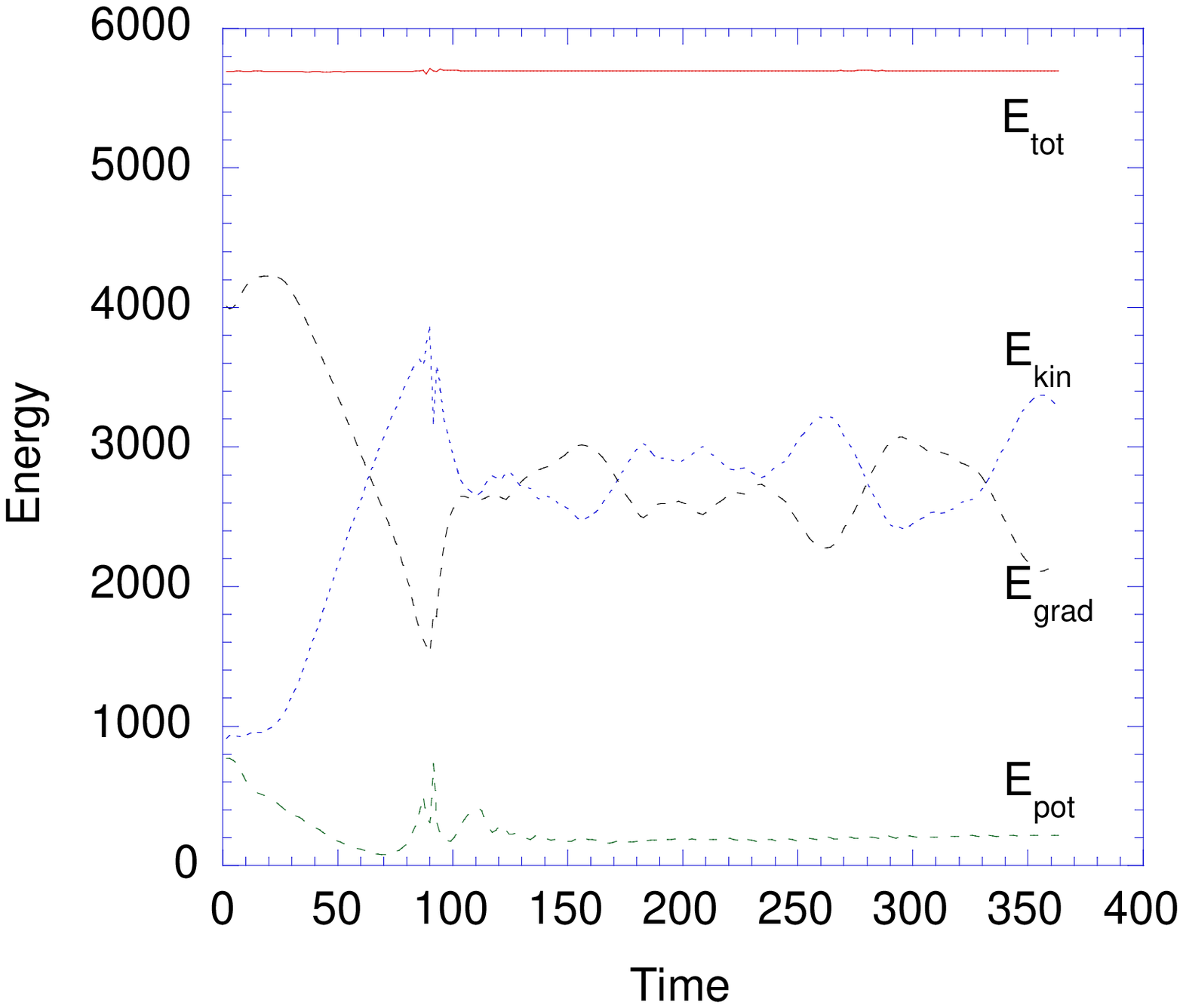,height=3.0in,bbllx=0in,bblly=2in,bburx=9in,bbury=8in}
\caption{\small{Energy of non-intersecting loop.  Shown are total, gradient, 
kinetic, and potential energy as a function of time for $\alpha = 0.01,
\psi = 0$, and $\lambda = 0.0625$.}}
\end{figure}
reduction of $N_{ax}$ of about $25 \%$.  Figure 6 depicts the energy of the 
collapsing loop. Clearly, the total energy is well conserved. A few percent 
of the loop energy is dissipated as massive radiation, shown here as 
$E_{\rm pot}$. 

\subsection{Conclusions}

In our 3D simulations of rotating loops, $r \simeq 0.75$ for 
$\ln(L/\delta) \simeq 4$.  We take $L=2R$.  In our 2D simulations of 
circular loops, $r \simeq 0.8$ for $\ln(L/\delta) \simeq 6$.  The earlier 
3D simulations of circular loops by two of us found 
$r \simeq 0.8$ for $\ln(L/\delta) \simeq 4$.  We conclude that $r$ is of 
order one when $\ln(L/\delta) \simeq 6$, and that $r$ does not appear to 
change when $\ln(L/\delta)$ is increased.  Hence we find the string decay 
contribution to the axion cosmological energy density Eq.~(\ref{3.28})] to 
be of the same order of magnitude as that from vacuum realignment 
[Eqs.~(\ref{3.18},\ref{3.19})].

\section{Wall decay}

During the QCD phase transition, when the axion acquires mass, each axion
string becomes the boundary of $N$ domain walls.  If $N>1$ there is a 
domain wall problem as mentioned in the introduction.  We assume $N=1$
here.  Each string is therefore attached to one domain wall.  The domain 
walls bounded by string are like pancakes or long surfaces with holes.  A 
simple statistical argument \cite{axwall} shows that domain walls which are 
not bounded by string, and therefore close onto themselves like spheres or 
donuts, are very rare.

When the axion acquires mass, the model of Eq.~(\ref{n26}) gets modified as 
follows:
\begin{equation}
{\cal L} = \frac{1}{2} \partial_{\mu} \phi_1 \partial^{\mu} \phi_1 +
\frac{1}{2}\partial_{\mu} \phi_2 \partial^{\mu} \phi_2 -\frac{\lambda}{4}
(\phi_1^2+\phi_2^2 -v_a^2)^2 + \eta v_a\phi_1
\label{vi1}
\end{equation}
The axion mass $m_a=\sqrt{\eta}$ for $v_a\gg m_a$.  We may set 
$\phi(x) = v_a e^{\frac{i}{v_a} a(x)}$ when restricting ourselves to
low energy configurations. The corresponding effective Lagrangian is :
\begin{equation}
{\cal L}_a = \frac{1}{2}\partial_\mu a\partial^\mu a + m_a^2
v_a^2\left[\cos
\left(\frac{a}{v_a}\right)-1\right] .
\label{vi2}
\end{equation}
Its equation of motion has the well-known Sine-Gordon soliton solution :
\begin{equation}
\frac{a(y)}{v_a}= 2\pi n + 4 \tan^{-1}\exp(m_a y)
\label{2.5}
\end{equation}
where $y$ is the coordinate perpendicular to the wall and $n$ is any
integer.  Eq.~(\ref{2.5}) describes a domain wall which interpolates
between neighboring vacua in the low energy effective theory (\ref{vi2}).
In the original theory (\ref{vi1}), the domain wall interpolates between
the unique vacuum back to that same vacuum going around the bottom of the
Mexican hat potential
$V(\phi^\dagger\phi)=\frac{\lambda}{4}(\phi^\dagger\phi - v_a^2)^2$ once.

The thickness of the wall is of order $m_a^{-1}$. The wall energy per unit
surface is $\sigma= 8 m_a v_a^2$ in the toy model of Eq.~(\ref{vi1}).  In
reality the structure of axion domain walls is more complicated than in the 
toy model, because not only the axion field but also the neutral pion field 
is spatially varying inside the wall \cite{Huang}. When this is taken into
account, the (zero temperature) wall energy/surface is found to be:
\begin{equation}
\sigma \simeq 4.2~f_\pi m_\pi f_a \simeq 9~m_a f_a^2\
\label{2.6}
\end{equation}
with $f_a \equiv v_a/N$.  For $N=1$, $f_a$ and $v_a$ are the same.
However, written in terms of $f_a$, Eq. (\ref{2.6}) is valid for $N \neq 1$
also.

Walls bounded by strings are of course topologically unstable.  In 
ref.\cite{axwall} we discussed three decay mechanisms - emission of 
gravitational waves, friction on the surrounding plasma and emission 
of axions - and concluded that decay into non-relativistic axions 
is very likely the dominant mechanism.  This is consistent with the 
computer simulations described below.  In the simulations the walls 
decay immediately, i.e. in a time of order their size divided by the 
speed of light, and the average energy of the radiated axions is
$\langle \omega_a \rangle\sim 7 m_a$. 

Let $t_3$ be the time when the decay effectively takes place in the 
early universe and let $\gamma \equiv \frac{\langle \omega_a
\rangle}{m_a(t_3)}$ be the average Lorentz $\gamma$ factor then of the 
axions produced.  The density of walls at time $t_1$ was estimated
\cite{axwall} to be of order $0.7$ per horizon volume.  Hence the 
average energy density in walls between $t_1$ and $t_3$ is of order
\begin{equation}
\rho_{\rm d.w.}(t)\sim (0.7)(9)m_a(t)\frac{f_a^2}{t_1}\left(
\frac{R_1}{R}\right)^3 \, .
\label{3.41n}
\end{equation}
We used Eq.~(\ref{2.6}) and assumed that the energy in walls simply 
scales as $m_a(t)$.  After time $t_3$, the number density of axions 
produced in the decay of walls bounded by strings is of order
\begin{equation}
n^{\rm d.w.}_a(t)\sim \frac{\rho_{\rm d.w.}(t_3)}{\langle \omega_a
\rangle} \left(\frac{R_3}{R}\right)^3\sim \frac{6}{\gamma}
\frac{f_a^2}{t_1}\left(\frac{R_1}{R}\right)^3 \, .
\label{3.42n}
\end{equation}
Note that the dependence on $t_3$ drops out of our estimate of
$n^{\rm d.w.}_a$.  

\subsection{Computer simulations}
\begin{figure}[b]
\psfig{figure=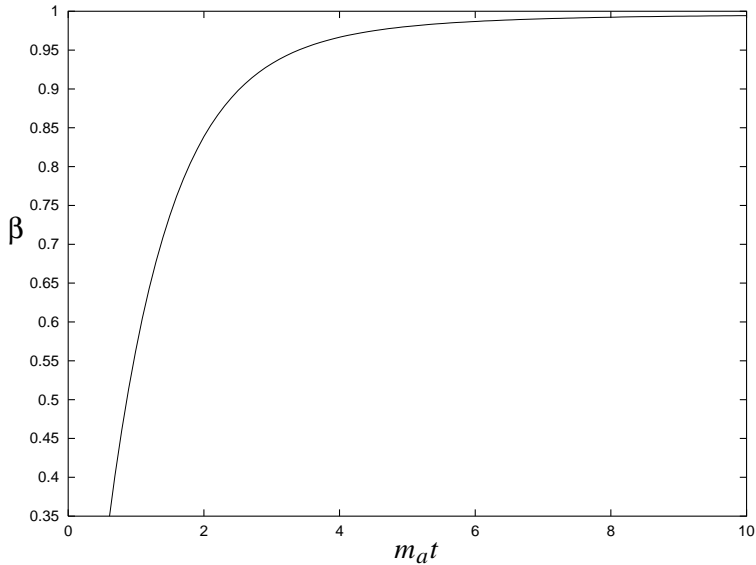,height=3.0in}
\caption{\small{Speed of the string core as a function of time for the
case $1/m_a=400$, $\sqrt{\lambda}/m_a =10$, and $v =0$.}}
\label{lorentz}
\end{figure}
We have carried out an extensive program of 2D numerical simulations of
domain walls bounded by strings using the Lagrangian of Eq.~(\ref{vi1}) in 
finite difference form.  We report our results in units where $v_a = 1$ 
and the lattice constant is the unit of length.  Thus the wall thickness 
is $1/m_a = 1/\sqrt{\eta}$ and the core size is $\delta = 1/\sqrt{\lambda}$.  
In the continuum limit, the dynamics depends upon a single critical 
parameter, $m_a \delta = m_a/\sqrt{\lambda}$.  Large two-dimensional grids 
$(\sim 4000\times 4000)$ were initialized with a straight domain wall 
initially at rest or with angular momentum. The initial domain wall was 
obtained by overrelaxation starting with the Sine-Gordon ansatz
$\phi_1+i\phi_2={\rm exp}(i\,{\rm tan^{-1}exp}(m_a y))$
inside a strip of length $D$ between the string and anti-string and the
true vacuum $(\phi_1=1,\,\phi_2=0)$ outside.  The string and anti-string
core were approximated by $\phi_1+i\phi_2=-{\rm tanh}(.583\,r/\delta)\,
{\rm exp}(\mp i\theta )$ where $(r,\theta)$ are polar coordinates about
the core center, and held fixed during relaxation. Stable domain walls
were obtained for $1/(\delta m_a )\gtwid 3$.  A first-order in time and
second order in space algorithm was used for evolution with a time step
$dt=0.2$. The boundary conditions were periodic and the total energy was
conserved to better than 1\%.  If the angular momentum was nonzero, the 
initial time derivative $\dot{\phi}$ was obtained by a finite difference
over a small time step.

\begin{figure}[t]
\psfig{figure=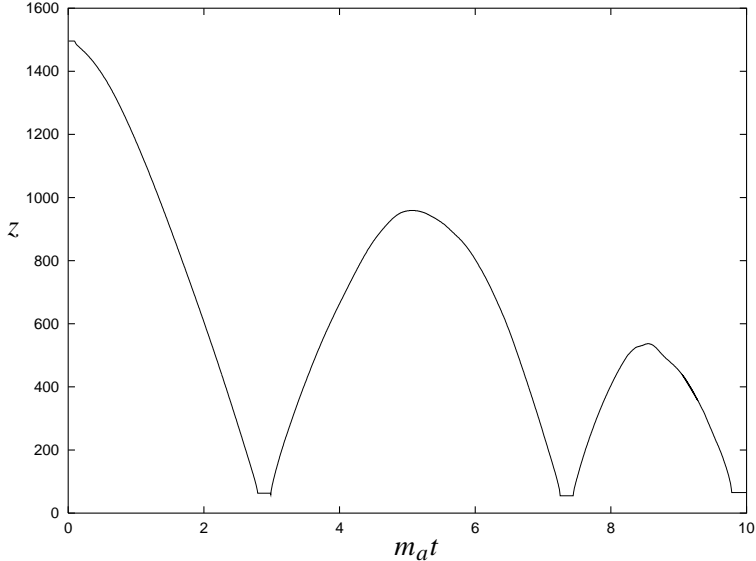,height=3.0in}
\caption{\small{Position of the string or anti-string core as a function
of time for $1/m_a=1000$, $\lambda = 0.0002, D=2896$, and $v=0$.
The string and anti-string cores have opposite $z$. They go through each
other and oscillate with decreasing amplitude.}}  
\label{bounce} 
\end{figure}

The evolution of the domain wall was studied for various values
of $\sqrt{\lambda}/m_a$, the initial wall length $D$ and the initial
velocity $v$ of the strings in the direction transverse to the wall.
Fig.~\ref{lorentz} shows the longitudinal velocity of the core as
a function of time for the case $m_a^{-1} = 400, \sqrt{\lambda}/m_a = 10$  
and $v=0$.   An important feature is the Lorentz contraction of the core.
For reduced core sizes $\delta/\gamma_s \ltwid  5$, where $\gamma_s$ is
the  Lorentz factor associated with the speed of the string core, there is
'scraping' of the core on the lattice accompanied by emission of spurious
high frequency radiation. This artificial friction eventually balances the
wall tension and leads to a terminal velocity. In our simulations we
always ensured being in the continuum limit.

\begin{figure}[t]
\psfig{figure=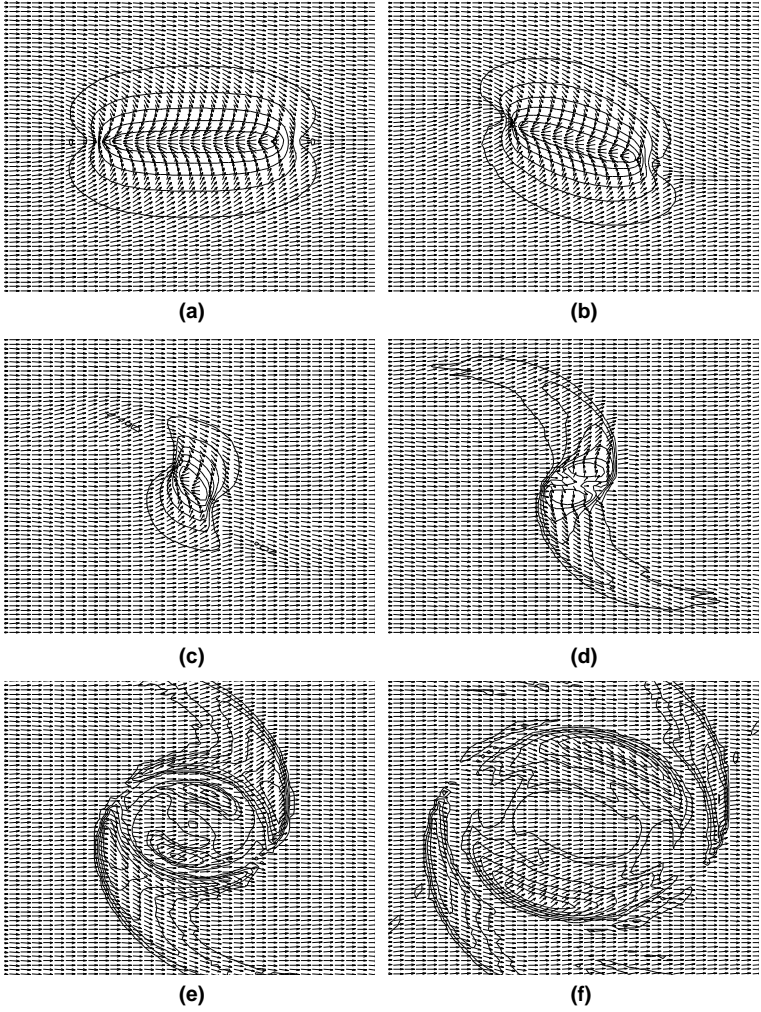,height=5.3in}
\caption{\small{Decay of a wall at successive time intervals $\Delta t =
1.2/m_a$ for the case $m_a^{-1}=100$, $\lambda = 0.01$, $D=524$, and $v =
0.6$.}}  
\label{rot}
\end{figure}
For a domain wall without rotation ($v=0$), the string cores strike 
head-on and go through each other.  Several oscillations of decreasing
magnitude generally occur before annihilation.  For $\gamma_s \simeq 1$
the string and anti-string coalesce and annihilate one another.  For
$\gamma_s \gtwid 2$, the strings go through each other and regenerate
another wall of reduced length.  The relative oscillation amplitude
decreases with decreasing collision velocity.  Fig.~\ref{bounce} shows
the core position as a function of time for the case
$m_a^{-1} = 1000, \sqrt{\lambda}/m_a = 14.3, D = 2896$.

We also investigated the more generic case of a domain wall with angular   
momentum.  The strings are similarly accelerated by the wall but string
and anti-string cores miss each other.  The field is displayed in
Fig.~\ref{rot} at six time steps for the case $m_a^{-1} = 100, \lambda =
0.01, D = 524 $ and $v = 0.6$.  No oscillation is observed.  All energy is 
converted into axion radiation during a single collapse.

\begin{figure}[t]
\psfig{figure=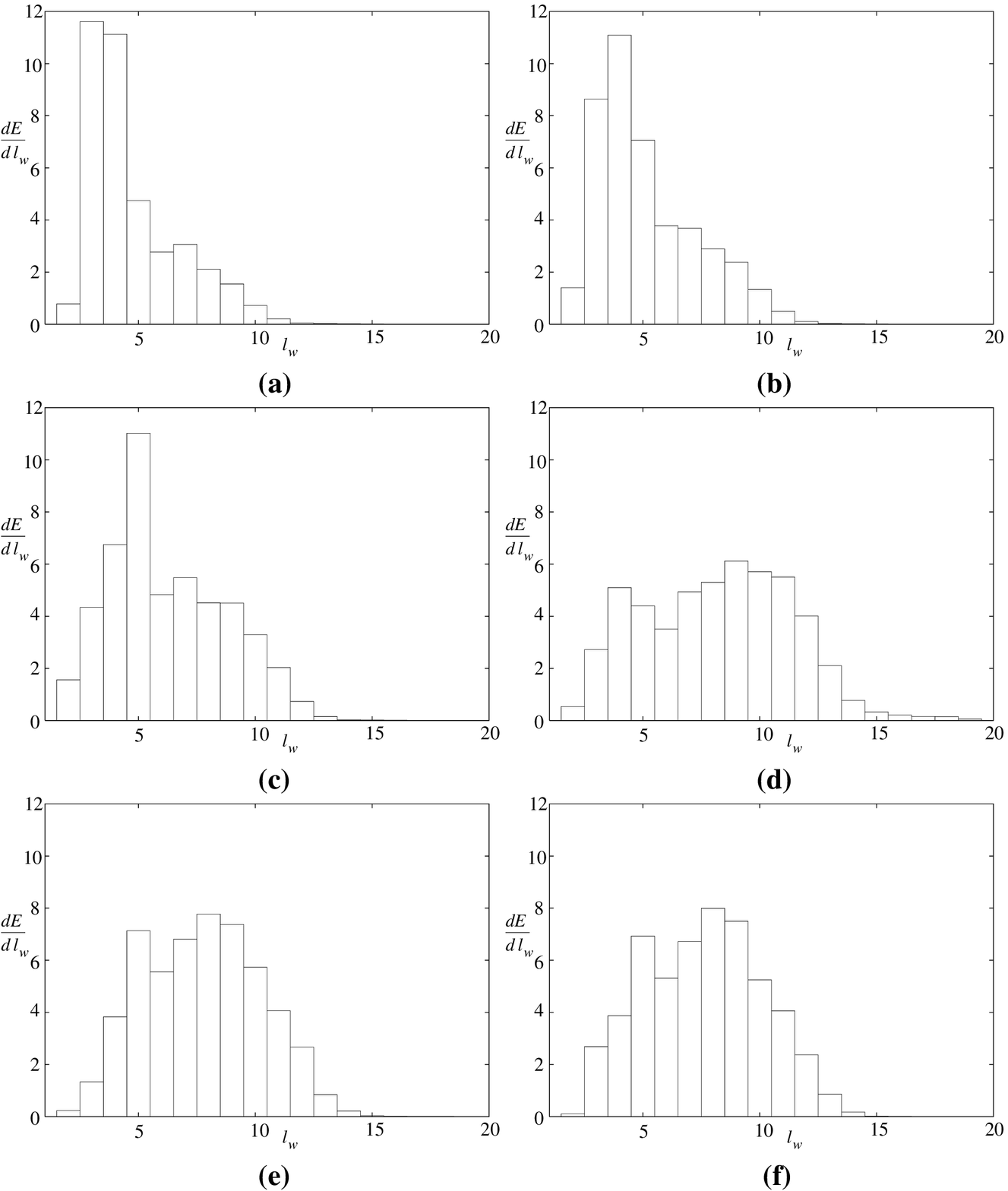,height=5.1in}
\caption{\small{Energy spectrum at successive time intervals $\Delta t =
1/m_a$, with $w_{\rm max}= \sqrt{8+m_a^2}$ and
$l_W=18 \ln(w/m_a)/\ln(w_{\rm max}/m_a)$ $+2$,
for the case $m_a^{-1}=500$, $\lambda=0.0032$,
$D=2096$ and $v=0.25$.}}
\label{spec}   
\end{figure}
We performed spectrum analysis of the energy stored in the $\phi$ field
as a function of time using standard Fourier techniques.  The
two-dimensional Fourier transform is defined by
\begin{equation}
\tilde{f}(\vec{p})\!=\!\frac{1}{\sqrt{V}}\!\sum_{\vec{n}}\!
\exp\!\left[2i\pi\left(\frac{p_xn_x}{L_x}+\frac{p_yn_y}{L_y}\right)\right]\!
f(\vec{n})
\end{equation}
where $V\equiv L_x L_y$.  The dispersion law is:
\begin{equation}
\omega_p=\sqrt{2 \left(2- \cos\frac{2\pi p_x}{L_x}- \cos\frac{2\pi
p_y}{L_y}\right)+m_a^2}.
\end{equation}
Fig.~\ref{spec} shows the power spectrum $dE/d\ln \omega$ of the $\phi$
field at various times during the decay of a rotating domain wall for the
case $m_a^{-1} = 500, \lambda = 0.0032, D = 2096$, and $v = 0.25$.
Initially, the spectrum is dominated by small wavevectors, $k\sim m_a$.  
Such a spectrum is characteristic of the domain wall.  As the wall
accelerates the string, the spectrum hardens until it becomes roughly
$1/k$ with a long wavelength cutoff of order the wall thickness $1/m_a$
and a short wavelength cutoff of order the reduced core size
$\delta/\gamma_s$.  Such a spectrum is characteristic of the moving
string.  
\begin{figure}[t]
\psfig{figure=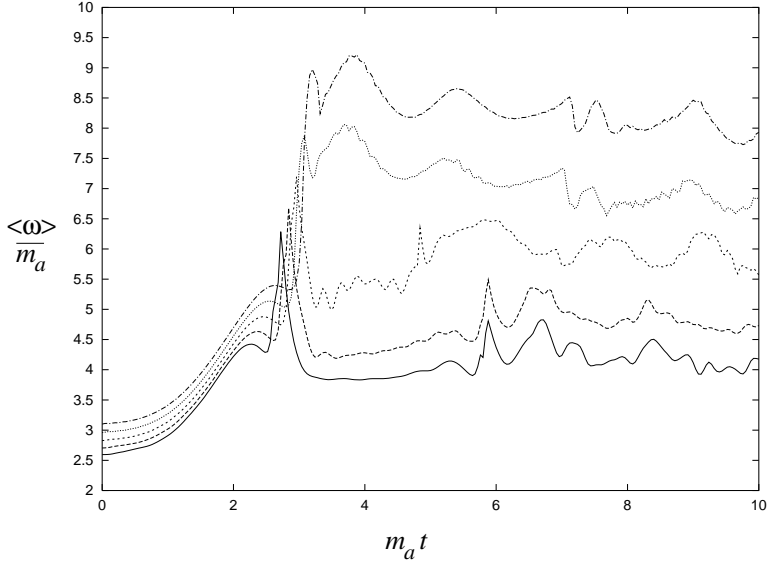,height=3.0in}
\caption{\small{$\langle \omega \rangle$ as a function of time for
$1/m_a=500$, $D=2096$, $v=0.25$ and $\lambda=$ 0.0004 (solid), 0.0008
(long dash), 0.0016 (short dash), 0.0032 (dot) and 0.0064 (dot dash).
After the wall has decayed, $\langle \omega \rangle$ is the average energy
of radiated axions.}}
\label{energy}   
\end{figure}   

\begin{figure}[t]
\psfig{figure=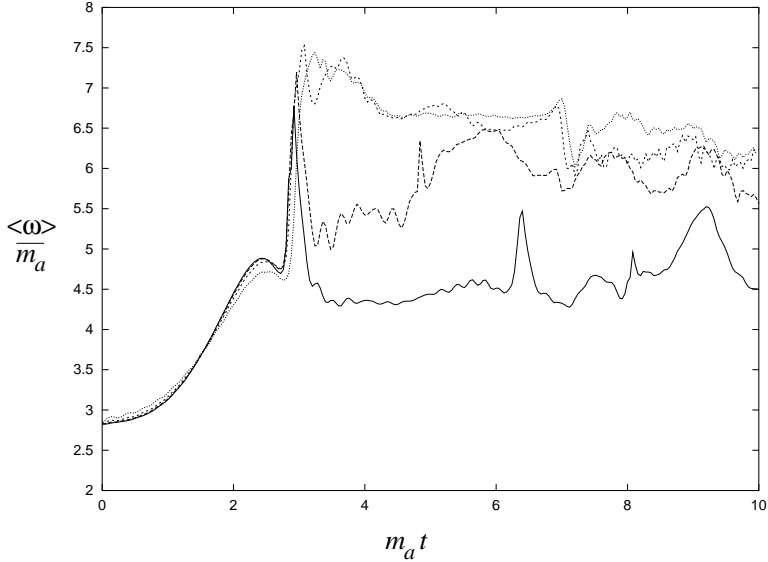,height=3.0in,bbllx=0in}
\caption{\small{Same as in Fig.~\ref{energy} except $\lambda = 0.0016$,
and $v=$ 0.15 (solid), 0.25 (long dash), 0.4 (short dash) and 0.6 (dot).}}
\label{energy2}
\end{figure}

Fig.~\ref{energy} shows the time evolution of $\langle \omega \rangle/m_a$
for $m_a^{-1} = 500, v = 0.25, D = 2096$ and  various values of
$\lambda$.  By definition,
\begin{equation}
\langle \omega \rangle = \sum_{\vec{p}} E_{\vec{p}}/
\sum_{\vec{p}} \frac{E_{\vec{p}}}{\omega_{\vec{p}}}
\end{equation}
where $E_{\vec{p}}$ is the gradient and kinetic energy stored in mode
$\vec{p}$ of the field $\phi$.  After the domain wall has decayed into
axions, $\langle \omega \rangle = \langle \omega_a \rangle$.
Fig.~\ref{energy2} shows the time evolution of $\langle \omega
\rangle/m_a$ for $m_a^{-1} = 500, \lambda = 0.0016, D = 2096$ and 
various values of $v$.

When the angular momentum is low and the core size is big, the strings
have one or more oscillations.  In that case,
$\langle \omega_a \rangle/m_a \simeq 4$,  which is consistent with
Ref.~\cite{Nagasawa}.  We believe this regime to be less relevant to wall
decay in the early universe because it seems unlikely that the angular
momentum of a wall at the QCD epoch could be small enough for the strings
to oscillate.

In the more generic case when no oscillations occur,
$\langle \omega_a \rangle/m_a \simeq 7$.  Moreover, we find that
$\langle \omega_a \rangle/m_a$ depends upon the critical parameter
$\sqrt{\lambda}/m_a$, increasing approximately as the logarithm of that
quantity; see Fig.~\ref{energy}.  This is consistent with the time 
evolution of the energy spectrum, described in Fig.~10.  For a domain wall,
$\langle \omega \rangle \sim m_a$ whereas for a moving string
$\langle \omega \rangle \sim m_a \ln (\sqrt{\lambda} v_a \gamma_s/m_a)$. 
Since we find the decay to proceed in two steps: 1) the wall energy is
converted into string kinetic energy, and 2) the strings annihilate
without qualitative change in the spectrum, the average energy of
radiated axions is
$\langle \omega_a \rangle \sim m_a \ln (\sqrt{\lambda} v_a/m_a)$. 
Assuming this is a correct description of the decay process for
$\sqrt{\lambda} v_a/m_a \sim 10^{26}$, then
$\langle \omega_a \rangle/m_a \sim 60$ in axion models of interest.

\subsection{Conclusions}

If we use the value $\gamma=7$ observed in our computer simulations, 
Eq.~(\ref{3.42n}) becomes
\begin{equation}
n^{\rm d.w.}_a(t)\sim\frac{f_a^2}{t_1}\left(\frac{R_1}{R}\right)^3 \, .
\label{3.43n}  
\end{equation}
In that case the contributions from vacuum realignment [Eq.~(\ref{3.18})
and Eq.~(\ref{3.19})] and domain wall decay are of the same order of
magnitude.  However, we find that $\gamma$ rises approximately linearly
with ln$(\sqrt{\lambda} v_a/m_a)$ over the range of ln$(\sqrt{\lambda}
v_a/m_a)$ investigated in our numerical simulations.  If this behaviour
is extrapolated all the way to ln$(\sqrt{\lambda} v_a/m_a) \simeq 60$,
which is the value in axion models of interest, then $\gamma \simeq 60$.
In that case the contribution from wall decay is subdominant relative to
that from vacuum realignment.

\section{The cold axion cosmological energy density}

For case 1, adding the three contributions, we estimate the present 
cosmological energy density in cold axions to be:
\begin{equation}
\rho_a (t_0) \sim 3 {f_a^2\over t_1} \left({R_1\over R_0}\right)^3 m_a
~~~~~~~~~(case~1)\ .
\label{3.38}
\end{equation}
Following \cite{vacmis}, we may determine the ratio of scale factors   
$R_1/R_0$ by assuming the conservation of entropy from time $t_1$ till
the present.  The number of effective thermal degrees of freedom at time
$t_1$ is approximately ${\cal N}_1 \simeq 61.75$.  We do not include
axions in this number because they are decoupled by then.  Let $t_4$ be a
time (say $T_4 = 4$ MeV) after the pions and muons annihilated but before
neutrinos decoupled.  The number of effective thermal degrees of freedom
at time $t_4$ is ${\cal N}_4 = 10.75$ with electrons, photons and three
species of neutrinos contributing.  Conservation of entropy implies
${\cal N}_1 T_1^3 R_1^3 = {\cal N}_4 T_4^3 R_4^3$.  The neutrinos decouple
before $e^+e^-$ annihilation.  Therefore, as is well known, the present
temperature $T_{\gamma 0} = 2.735~^0$K of the cosmic microwave background
is related to $T_4$ by:  
${11\over 2} T_4^3 R_4^3= 2~T_{0\gamma}^3~R_0^3$.  Putting everything 
together we have 
\begin{equation}
\left({R_1\over R_0}\right)^3 \simeq 0.063 \left({T_{\gamma, 0}\over T_1}
\right)^3\ .
\label{3.39}
\end{equation}
Combining Eqs.~(\ref{3.14}), (\ref{3.38}) and (\ref{3.39}),
\begin{equation}
\rho_a (t_0) \sim 10^{-29} {\mbox{gr}\over \mbox{cm}^3}  
\left({f_a\over 10^{12} \mbox{GeV}}\right)^{7/6}~~~~~(case~1)\ .
\label{3.40}
\end{equation}  
Dividing by the critical density $\rho_c = {3H_0^2\over 8\pi G}$, we find:
\begin{equation}
\Omega_a \sim \left({f_a\over 10^{12} \mbox{GeV}}\right)^{7/6}
\left({0.7\over h}\right)^2 ~~~~~~~~~~(case~1)
\label{3.41}
\end{equation}  
where $h$ is defined as usual by $H_0 = h~100 $km/s$\cdot$Mpc.

For case 2, we have
\begin{equation}
\Omega_a \sim {1 \over 6}\left({f_a\over 10^{12} \mbox{GeV}}\right)^{7/6}
\left({0.7\over h}\right)^2 \alpha^2(t_1)~~~~~~~~(case~2)
\label{3.42}
\end{equation}  
where $\alpha(t_1)$ is the unknown misalignment angle.

\section{Pop.~I and Pop.~II axions}

On the basis of the discussion in sections 2 to 4 we distinguish two 
kinds of cold axions:\\

I) axions which were produced by vacuum realignment or string decay and
which were not hit by moving domain walls. They have typical momentum
$\langle p_I(t_1)\rangle \sim \frac{1}{t_1}$ at time $t_1$ because they
are associated with axion field configurations which are inhomogeneous on
the horizon scale at that time. Their velocity dispersion is of order:
\begin{equation}
\beta_I(t)\sim \frac{1}{m_at_1}\left(\frac{R_1}{R}\right)\simeq 3\cdot
10^{-17}\left(\frac{10^{-5}{\rm eV}}{m_a}\right)^{5/6}\frac{R_0}{R} \ .  
\label{3.48}
\end{equation}
We call these axions population I.\\

II) axions which were produced in the decay of domain walls. They have
typical momentum $\langle p_{II}(t_3)\rangle \sim \gamma m_a(t_3)$ at time
$t_3$ when the walls effectively decay. Their velocity dispersion is of
order:
\begin{equation}
\beta_{II}(t)\sim \gamma \frac{m_a(t_3)}{m_a}\frac{R_3}{R}\simeq 10^{-13}
\left(\ \gamma \frac{m_a(t_3)}{m_a}\frac{R_3}{R_1}\right)
\left(\frac{10^{-5}{\rm eV}}{m_a}\right)^{1/6} \frac{R_0}{R} \, .
\label{3.49}  
\end{equation}
The factor $q \equiv \gamma \frac{m_a(t_3)}{m_a}\frac{R_3}{R_1}$
parametrizes our ignorance of the wall decay process.  We expect $q$ to
be of order one but with very large uncertainties.  There is however a
lower bound \cite{axwall}: 
$q \gtwid \frac{\gamma}{130}\left(\frac{10^{-5}{\rm eV}}{m_a}\right)^{2/3}$.
Since our computer simulations suggest $\gamma$ is in the range 7 to 60,
the axions of the second population (pop. II) have much larger velocity   
dispersion than the pop. I axions.\\

Note that there are axions which were produced by vacuum realignment or
string decay but were hit by relativistically moving walls at some
time between $t_1$ and $t_3$.  These axions are relativistic just after
getting hit and therefore are part of pop.~II rather than pop.~I.

The fact that there are two populations of cold axions, with widely
differing velocity dispersion, has interesting implications for the 
formation and evolution of axion miniclusters \cite{Hogan,Kolb}.  We 
showed \cite{axwall} that the QCD horizon scale density perturbations 
in pop.~II axions get erased by free streaming before the time $t_{eq}$
of equality between radiation and matter whereas those in pop.~I axions
do not.  Hence, whereas pop.~I axions form miniclusters, pop.~II axions
form an unclustered component of the axion cosmological energy density 
which guarantees that the signal in direct searches for axion dark 
matter is on all the time.

The primordial velocity dispersion of pop.~II axions may some day be 
measured in a cavity-type axion dark matter detector \cite{experim}.  If 
a signal is found in such a detector, the energy spectrum of dark matter 
axions will be measured with great resolution.  It has been pointed out 
that there are peaks in the spectrum \cite{peaks} because late infall 
produces distinct flows, each with a characteristic local velocity vector. 
These peaks are broadened by the primordial velocity dispersion, given in 
Eqs.~(\ref{3.48}) and (\ref{3.49}) for pop.~I and pop.~II axions respectively.  
These equations give the velocity dispersion in intergalactic space.  When 
the axions fall onto the galaxy their density increases by a factor of order 
$10^3$ and hence, by Liouville's theorem, their velocity dispersion increases 
by a factor of order 10.  [Note that this increase is not isotropic in 
velocity space.  Typically the velocity dispersion is reduced in the
direction longitudinal to the flow in the rest frame of the galaxy whereas
it is increased in the two transverse directions.]  The energy dispersion 
measured on Earth is $\Delta E = m_a\beta \Delta \beta$ where $\beta
\simeq 10^{-3}$ is the flow velocity in the rest frame of the Earth.
Hence we find $\Delta E_I\sim 3\cdot 10^{-19}\left(\frac{10^{-5}{\rm
eV}}{m_a}\right)^{5/6} m_a$ for pop.~I axions and
$\Delta E_{II}\sim 10^{-15}~q~
\left(\frac{10^{-5}{\rm eV}}{m_a}\right)^{1/6}m_a$ for pop.~II.  The
minimum time required to measure $\Delta E$ is $(\Delta E)^{-1}$.  This
assumes ideal measurements and also that all sources of jitter in the
signal not due to primordial velocity dispersion can be understood.  There
is little hope of measuring the primordial velocity dispersion of pop.~I
axions since
$(\Delta E_I)^{-1}\sim 10~{\rm years} \left(\frac{10^{-5}{\rm eV}}{m_a}
\right)^{1/6}$.  However
$(\Delta E_{II})^{-1} \sim ~$day $q^{-1} \left(\frac{10^{-5}{\rm eV}}{m_a}   
\right)^{5/6}$, and hence it is conceivable that the primordial velocity
dispersion of pop.~II axions will be measured.

\section*{Acknowledgements}

This research was supported in part by DOE grant DE-FG05-86ER40272 at the 
University of Florida and by DOE grant W-7405-ENG-048 at Lawrence Livermore 
National Laboratory.

\end{document}